\documentclass[runningheads]{llncs}
\usepackage{graphicx}
\usepackage{booktabs}
\usepackage{hyperref}
\usepackage{tabularx}
\usepackage[inkscapeformat=pdf]{svg}
\usepackage{url}

\begin{document}
\title{Chances and Challenges of the \\Model Context Protocol in \\Digital Forensics and Incident Response}

\titlerunning{MCP in Digital Forensics}
\authorrunning{Hilgert et al.}

\author{Jan-Niclas Hilgert\and
Carlo Jakobs\and
Michael Külper\and\\
Martin Lambertz\and
Axel Mahr\and
Elmar Padilla}
%
\institute{Fraunhofer FKIE, Zanderstr. 5, 53177 Bonn, Germany}
\maketitle              

%
\begin{abstract}
Large language models hold considerable promise for supporting forensic investigations, but their widespread adoption is hindered by a lack of transparency, explainability, and reproducibility. 
This paper explores how the emerging Model Context Protocol can address these challenges and support the meaningful use of LLMs in digital forensics. 
Through a theoretical analysis, we examine how MCP can be integrated across various forensic scenarios---ranging from artifact analysis to the generation of interpretable reports. 
We also outline both technical and conceptual considerations for deploying an MCP server in forensic environments.
Our analysis reveals a wide range of use cases in which MCP not only strengthens existing forensic workflows but also facilitates the application of LLMs to areas of forensics where their use was previously limited.
Furthermore, we introduce the concept of the inference constraint level---a way of characterizing how specific MCP design choices can deliberately constrain model behavior, thereby enhancing both auditability and traceability.
Our insights demonstrate that MCP has significant potential as a foundational component for developing LLM-assisted forensic workflows that are not only more transparent, reproducible, and legally defensible, but also represent a step toward increased automation in digital forensic analysis. However, we also highlight potential challenges that the adoption of MCP may pose for digital forensics in the future.

\end{abstract}

%
\section{Introduction}

Over the past few years, large language models (LLMs) have emerged as a breakthrough in artificial intelligence, demonstrating impressive capabilities in language understanding, reasoning, and tool use. Their rapid evolution and integration have prompted researchers and practitioners to explore their potential well beyond traditional Natural Language Processing tasks---including applications in digital forensics and incident response. Several studies highlight AI’s growing role in digital forensics, with some demonstrating its potential to automate evidence analysis~\cite{costantini2019digital}, and others exploring explainable AI to enhance the transparency and reliability of forensic triage and artifact extraction~\cite{hall2022explainable}. An expanding line of research~\cite{jarrett2021impact,tynan2024integration} underscores the increasing interest in leveraging LLMs and other AI methods for digital forensic applications including artifact identification, anomaly detection, script generation, incident analysis, and educational support in forensic training scenarios. LLMs are considered promising in digital forensics due to their capacity to interpret natural language prompts, simulate procedural reasoning, and assist with both technical and explanatory tasks across investigative contexts~\cite{scanlon2023chatgpt}. Wickramasekara et al.~\cite{wickramasekara2025exploring} argue that LLMs are increasingly considered in digital forensics to help address investigation backlogs by enhancing efficiency, improving traceability, and lowering technical and legal barriers across multiple phases of the forensic process.


Despite their potential, LLMs currently face several significant barriers to autonomous use in digital forensics. Their “black box” nature raises critical concerns around transparency, accountability, and explainability---attributes essential to maintaining trust and evidentiary integrity in legal and investigative contexts. 
In addition, LLMs may hallucinate facts, misinterpret evidence, or produce incomplete or unverifiable reasoning chains, all of which undermine the reliability of their outputs. Furthermore, LLMs often lack domain-specific knowledge required to accurately interpret low-level forensic artifacts, making them prone to incorrect assumptions when working with unfamiliar data structures or tool outputs. Adding to these limitations, making relevant forensic data available to an LLM in a structured and usable form is itself a complex task.


In November 2024, Anthropic introduced the Model Context Protocol (MCP), a standardized framework designed to connect LLMs \textit{"to the systems where data lives"}\footnote{\url{https://www.anthropic.com/news/model-context-protocol}}. MCP enables LLMs to interact with external tools and data sources through so-called \textit{MCP servers}, which expose well-defined functions for tasks such as data retrieval or tool execution. Since its release, MCP has quickly gained recognition as a foundational component in the evolving AI ecosystem. Adoption by major organizations is accelerating this momentum: OpenAI has integrated MCP into its Agents SDK and has announced plans to support it in the Responses API\footnote{\url{https://openai.com/index/new-tools-and-features-in-the-responses-api/}} and the ChatGPT desktop application. Microsoft has introduced native MCP support in Windows 11\footnote{\url{https://blogs.windows.com/windowsexperience/2025/05/19/securing-the-model-context-protocol-building-a-safer-agentic-future-on-windows/}}, alongside a dedicated C\# SDK, enabling tighter integration with the Windows architecture and tools like Copilot Studio. Google DeepMind has also confirmed its intention to adopt MCP within its Gemini model ecosystem\footnote{\url{https://blog.google/technology/google-deepmind/google-gemini-updates-io-2025/\#performance}}, recognizing it as a rapidly emerging open standard for AI-agent connectivity. 

While MCP presents considerable potential, as a relatively recent innovation, it has received limited attention in the literature to date. The most comprehensive analysis so far is provided by Hou et al.~\cite{hou2025model} who examine the protocol’s architecture, security risks, and broader adoption in general AI contexts. While their work provides a foundational understanding of MCP at the systems level, it does not explore its potential in domain-specific settings such as digital forensics---where many of MCP’s strengths directly address the core challenges of applying LLMs, including limited tool access, lack of domain context, and the need for transparent and verifiable reasoning.

This paper addresses this gap by exploring the emerging role of MCP in digital forensics and incident response. We present several use cases demonstrating how MCP can support a variety of tasks, ranging from artifact analysis to the generation of comprehensible reports and outline important considerations for deploying MCP servers within forensic workflows. In addition to these aspects, we also highlight potential challenges that may arise as MCP becomes more widely adopted. By doing so, this work aims to raise awareness of both the benefits and limitations of MCP, and its potential impact on the future of digital forensics and incident response.


%
\section{Model Context Protocol}
MCP is an open standard designed to provide a unified interface for connecting LLM applications with external tools and resources. Unlike traditional integrations that depend on hardcoded API connections, MCP enables LLMs to independently locate, choose, and interact with services and data sources based on the given task context. It also incorporates human-in-the-loop capabilities, allowing users to contribute data or validate actions when necessary. The following sections outline the structure of MCP, based on its official specification~\cite{mcp-spec}.


\subsection{Architecture}
\begin{figure}
    \centering
    \includegraphics[width=\linewidth]{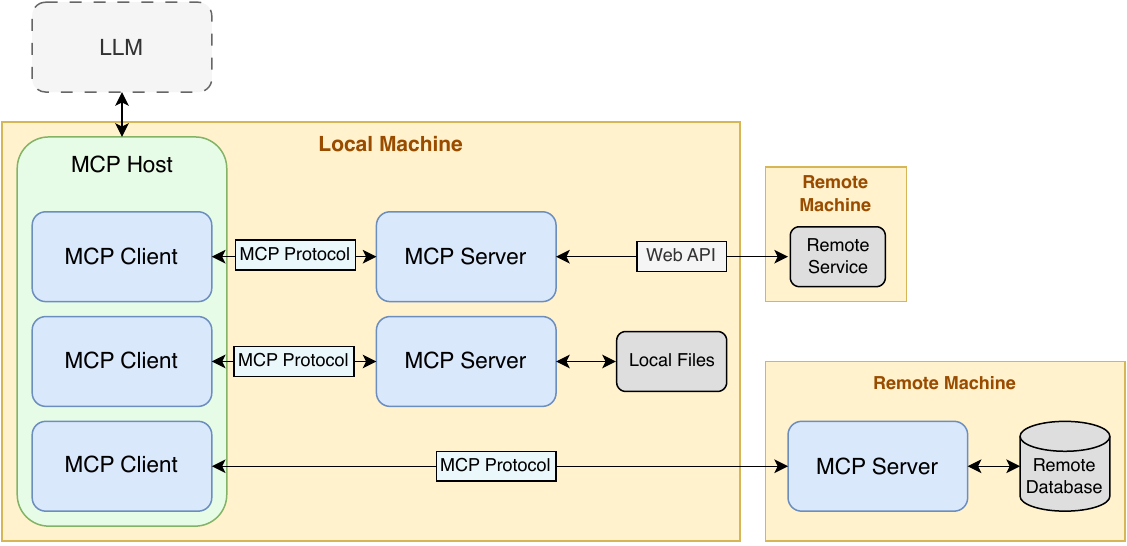}
    \caption{Overview of the MCP architecture, illustrating interactions with both local and remote resources. Depending on the specific host configuration, the LLM may also run locally.}
    \label{fig:mcp-architecture}
\end{figure}

MCP employs a client-server architecture, where one host application can establish connections with several MCP servers. Figure~\ref{fig:mcp-architecture} gives an exemplary overview. The architecture consists of the following core components:

\begin{itemize}
    \item \textbf{MCP Host:} The MCP host is the LLM-enabled application (e.g. VS Code, Claude Desktop) that seeks to interact with a resource. It separately coordinates the integration of the LLM and of one or more MCP clients.
    
    \item \textbf{MCP Client:} The MCP client is a component within the MCP host and sends requests to an MCP server. Each MCP client maintains a dedicated one-to-one connection with an MCP server. The MCP specification provides that a separate client-server pair is used for each type of resource. For example, one pair may be used to interface with MySQL databases, while another may handle interaction with IDA Pro.

    \item \textbf{MCP Server:} The MCP server is the main component of interest for practitioners and needs to be implemented for dedicated access to the corresponding resource. It connects to the actual resource, either locally (via direct access) or remotely (via a Web API), and exposes its functionalities to the corresponding MCP client.
\end{itemize}

\subsection{Capabilities}
In addition to the communicating parties, the MCP specification defines so-called \textit{capabilities} which constitute (partly mandatory) extensions to the raw message exchange between MCP client and server and can be configured by either side.

\begin{itemize}
    \item \textbf{Roots:} Defined by the MCP client, roots specify the files and directories that the MCP server is permitted to access. They are represented as URIs, which may point to local file paths, directories, or remote API endpoints.

    \item \textbf{Resources:} This capability represents the data sources accessed by the MCP server. These may include persistent data stored in files, such content from databases, images, or log files, as well as transient text or binary data handled by a connected service. Depending on the MCP client implementation, the client may autonomously select which resources to access within the defined roots.

    \item \textbf{Tools:} Defined by the MCP server, tools represent the specific functionalities it provides to the client---ranging from simple data retrieval to complex operations. Each tool is identified by a name and accompanied by a description of its functionality. 
    The choice of which tool to invoke is delegated to the LLM. 
    Furthermore, MCP supports \textit{dynamic tool discovery}, allowing clients to query available tools at any time, while enabling the server to update or modify tools at runtime.

    \item \textbf{Sampling:} When the MCP server requires additional information outside its scope of expertise, it can submit queries back to the LLM via the MCP client. These queries must be reviewed and approved by the user before being passed to the LLM. The LLM’s response is then returned to the server through the client, again subject to user review. 

    \item \textbf{Prompts:} This capability allows the MCP server to provide reusable prompt templates, which are exposed to the MCP client and can be selected by the user. These prompts may include resource-specific dynamic content and can be chained together to support entire workflows.

    \item \textbf{Logging:} Defined by the MCP server, this defines the log message level for log messages sent to the client and may let the client adjust it.

    \item \textbf{Experimental:} This capability serves as a wildcard for custom, application-specific features that may be implemented by either the MCP client or server.
\end{itemize}

\subsection{Procedure}
MCP uses JSON-RPC as the underlying protocol for message exchange between the client and server. Messages can be transmitted via stdin/stdout when the client and server run on the same system, or via HTTP POST for network-based communication. For HTTP-based interactions, MCP optionally supports OAuth~2.1 for authorization, to be implemented by both the client and server as needed.

MCP defines three primary message types: request, response and notification. All three message types have a predefined JSON-RPC format and can be initiated by both the client and the server. Notifications are one-way messages and used to signal events such as task cancellation or to report task progress.

The typical use of MCP is structured into three phases:
\begin{enumerate}
	\item \textbf{Initialization:} MCP client and server exchange and negotiate capabilities, and agree on a protocol version.

	\item \textbf{Operation:} The client and server exchange messages according to the negotiated capabilities. A typical interaction starts with a user prompt, which the LLM converts into a sequence of MCP requests containing capability-specific calls. These requests, which may be batched, are sent to the server. Depending on the capability and logging strategy, the server may trigger additional nested message exchanges and LLM interactions. The final result is returned to the MCP host.

	\item \textbf{Shutdown:} The connection is closed gracefully, with the shutdown procedure differing based on the chosen transport method (stdio or HTTP).
\end{enumerate}

\subsection{Current Implementations}
There are already thousands of MCP server implementations available, covering a wide range of use cases, including database integration, file system operations, software development and application automation, image and video processing, blockchain interaction, and more. A comprehensive and regularly updated list can be found on the Glama website\footnote{\url{https://glama.ai/mcp/servers}}.

To support the development of MCP-based systems, several codebases are available for implementing MCP clients and servers. Official SDKs provided by Anthropic include support for Python, TypeScript, C\#, Java, Kotlin, and Swift. In addition, community-driven projects offer implementations in other languages such as Go and Rust, as well as a widely adopted, feature-rich Python implementation known as \textit{fastmcp}\footnote{\url{https://github.com/jlowin/fastmcp}}.

%
\section{Use Cases in Digital Forensics and Incident Response}

AI, in general, and LLMs, in particular, offer significant potential to support forensic investigations. However, for such technologies to be adopted on a broad scale and especially admissible in court, several critical challenges must be addressed. One central point of criticism, which continues to foster skepticism, is the lack of transparency in the outputs generated by AI methods~\cite{garrett_rudin.2023.interpretable-forensics}. Even modern reasoning models, capable of providing a chain of thought, do not yet meet the reproducibility and explainability required for their unreserved use in investigative contexts. One contributing factor is that the way an LLM interprets data and formulates conclusions is often difficult for humans to comprehend. As the model is granted greater freedom to analyze information and generate its inferences, the transparency of its reasoning diminishes. This increased freedom makes it harder for humans to trace or validate the model's logic and increases the risk of misinterpretation and hallucinations. We refer to the extent of an LLM's independent reasoning as its \textit{inference freedom level}.

This shortcoming affects not only parties, such as judges or attorneys, but also the investigators themselves. Investigators and independent experts must be able to present their conclusions in a manner that is comprehensible and verifiable in a legal setting, which is not feasible if AI models are applied in a black-box manner~\cite{hall2022explainable,tynan2024integration}. An essential requirement in this context is auditability. The use of AI systems in forensic investigations must be accompanied by robust logging and documentation mechanisms that enable the reconstruction of each step in the analytical process. This documentation includes the input data, model versions, prompt formulations, and generated outputs. Only with such traceable and verifiable records can investigators and expert witnesses credibly justify their conclusions, and can courts evaluate the reliability and admissibility of AI-assisted findings.

From our perspective, MCP servers can contribute to resolving or at least mitigating many of the aforementioned issues. In particular, they enable an assessment of the AI methods employed, which in turn can be used to evaluate their admissibility in forensic investigations. For traditional forensic tools, a range of requirements must be met in order to conform to established best practices. Daniel and Daniel state that ``\textit{[f]or any tool to be forensically sound, it must be definable, predictable, repeatable, and verifiable}''~\cite{daniel.2012.df-for-legal-pros} and Horsman emphasizes the need for rigorous tool testing~\cite{horsman.2019.tool-testing}. MCP servers as building blocks for larger AI-based workflows can fulfill these requirements, as they can be assessed and tested separately from the LLM orchestrating them. They can significantly promote the traceability and comprehensibility of the results produced by an LLM, helping to reduce the black-box nature that poses significant challenges in forensic and legal contexts~\cite{garrett_rudin.2023.interpretable-forensics,wickramasekara2025exploring}. However, this does not occur automatically. Since an MCP server can essentially implement arbitrary functionalities, including querying other LLMs, it is important to assess a server's functionality concerning its suitability for forensic purposes.

We argue that MCP servers can be classified based on how much they allow the LLM discretion in interpreting the data returned. For example, to determine the group membership of a user account, an MCP server might return only the contents of the Windows registry in plain text to the controlling LLM. In this case, the LLM is responsible for correctly parsing the data and correlating values from the relevant registry keys. The MCP server merely provides the data in a format readable by the LLM, while all interpretation is left entirely to the LLM.

At the other end of the spectrum is an MCP server that directly provides the functionality to return the group membership of accounts. The corresponding logic, i.e., parsing the registry and linking the relevant keys, can be implemented within the MCP server in a transparent and auditable manner. In this scenario, the LLM has no interpretive discretion when answering whether a given account belongs to a specific group.
Naturally, intermediate levels also exist---for instance, when the Windows registry data has already been preprocessed or filtered.

\begin{figure}
    \centering
    \includegraphics[width=0.75\linewidth]{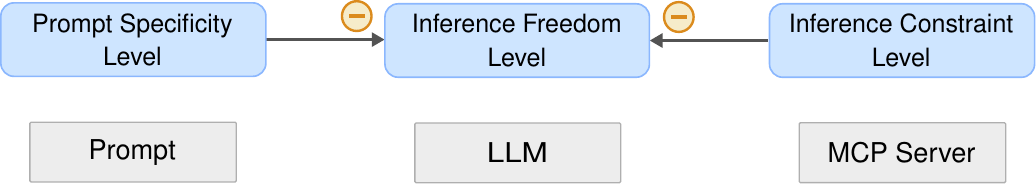}
    \caption{Relation between prompt specificity level, inference constraint level, and the inference freedom level.}
    \label{fig:no-pyramids}
\end{figure}

We call this spectrum the \textit{inference constraint level} of an MCP server. This level describes how much an MCP server constrains the inference freedom level. The inference constraint level of an MCP server is inversely proportional to the inference freedom level of an LLM. As constraint increases, the model's capacity for interpretive deviation decreases, ensuring greater alignment with forensic standards. Fig.~\ref{fig:no-pyramids} illustrates this concept on a schematic level.

Fig.~\ref{fig:no-pyramids} also illustrates that the inference freedom level is not limited solely by the inference constraint level; the prompt also influences it. The prompt may be highly specific, e.g., prescribing the execution of certain MCP functions in a defined order, or it may be abstract and generic. Accordingly, we define the \textit{prompt specificity level} as the variable that captures this dimension.

An MCP server's appropriate inference constraint level depends heavily on its intended use. For artifact analysis, a very high level is required. The MCP server should provide only transparent and verifiable functionality in such cases. In contrast, when an MCP server is used to generate guesses for password cracking, it may employ arbitrary or non-explainable methods, as the individual guesses are available for inspection afterward.

The following sections present various applications that benefit from integrating MCP servers to constrain the inference freedom level of LLMs. These use cases span digital forensic investigations, incident response scenarios, test data generation, and the development of knowledge bases.


%
\subsection{Knowledge Database}
One use case of an MCP server within forensic workflows is to enable an LLM to access and query knowledge databases. 
These may include scientific papers, technical specifications, documentation, or dedicated forensic knowledge bases such as SOLVE-IT~\cite{hargreavesSOLVEITProposedDigital2025}.
This approach also addresses a limitation noted by Scanlon et al.~\cite{scanlon2023chatgpt}, who observed that LLMs may lack information on recent artifacts due to the static nature of their training data. 
In contrast, MCP servers could query the internet or fall back to maintained data sources.

The MCP server can present the LLM with varying levels of structured data depending on the implementation.
In its simplest form, it supplies raw, unformatted content directly retrieved from the knowledge sources.
In more advanced setups, it can enrich this input with contextual metadata---such as source reliability, publication date, document type, or cross-references to related materials. 
This added structure helps the LLM reason more effectively about the information it receives.

By enabling the LLM to access structured forensic knowledge, the MCP server supports forensic processes in multiple ways.
At its simplest, it allows investigators to retrieve relevant information more easily and quickly by expressing their queries in natural language. 
Beyond that, the MCP server can help extract detailed guidance---for example, where to find a specific artifact, how to parse it, or how to interpret it. 
This capability can also be combined with other MCP servers focused on artifact analysis.

For example, consider an investigator analyzing a mobile device that contains an unfamiliar SQLite database. 
By querying the MCP server with a natural language question such as ``What does the msg\_store.db file from WhatsApp contain, and how can I extract message timestamps?'', the LLM can access structured knowledge from relevant documentation, prior case analyses, or published forensic research.

%
\subsection{Artifact Analysis}

Digital forensic investigations often involve analyzing data from application- or operating system-specific sources. Common examples include SQL or SQLite databases used by messaging apps and other mobile applications, though a wide variety of formats may be encountered. Before such data can be examined, it typically must be parsed into a structured, human-readable format. A common approach is to extract and convert the contents into plain text, which is then passed to an LLM for analysis. MCP simplifies this process by enabling LLMs to interact directly with databases or other data sources. For instance, official MCP servers for PostgreSQL and SQLite already support such direct access\footnote{\url{https://github.com/modelcontextprotocol/servers/}}.

While this approach simplifies data ingestion, it presents two key limitations: (i) it remains generic, requiring the LLM to interpret and analyze the raw data itself, and (ii) it often involves transmitting large volumes of data, constrained by the LLM’s context window. To overcome these issues, we propose the use of \textit{application-specific and context-aware MCP servers}. MCP can act as a \textit{semantic layer} between the LLM and data sources, e.g. including databases, log files, or application-specific formats, without requiring the model to directly parse or understand the underlying structure, thus reducing its inference freedom level. By exposing well-defined tools with meaningful names and descriptions, the MCP server allows the LLM to select appropriate operations based on what it is trying to accomplish in response to the user’s query. The MCP server then performs targeted, domain-specific tasks and returns only the relevant results. This not only increases the accuracy and efficiency of the analysis, but also significantly reduces context window usage by avoiding unnecessary data transfer.

\paragraph{Forensic Tool Use.}
One example of this approach is the use of an MCP server to enable LLMs to interact with arbitrary forensic applications and tools. In digital forensics, this opens the door for LLMs to use established forensic utilities rather than reproducing functionality that existing tools already implement. MCP can interface with anything from basic command-line tools like \texttt{file} or \texttt{strings} to advanced frameworks such as Volatility or The Sleuth Kit (TSK), and even GUI-based software---provided the MCP server manages the interaction appropriately.

However, as discussed before, the inference freedom level of an LLM when interacting with forensic tools depends heavily on how the corresponding MCP server is implemented. In a minimal setup, the MCP server may act as a thin wrapper that simply allows the LLM to execute a tool and return its raw output. While this approach can be effective when the prompt is explicit, it poses several challenges for more general or ambiguous instructions: the LLM may fail to recognize which tool is appropriate, lack the necessary knowledge to construct correct command-line arguments, or misinterpret the tool’s output.

To illustrate these limitations, consider the prompt:
\textit{“List all deleted files of the file system on disk.dd.”}


In this scenario, the LLM is granted access to a basic MCP server that exposes a tool called \texttt{run\_tsk\_command()}, allowing execution of arbitrary commands from TSK. This tool requires the LLM to specify the exact command and its parameters---such as invoking \texttt{fls} with the appropriate flags. As a result, the LLM must (i) identify that TSK is the appropriate toolkit, (ii) construct the correct command syntax, and (iii) correctly interpret the output. This setup leaves significant room for error. Without sufficient context, the LLM may hallucinate commands, misuse tool arguments, or misread results---ultimately compromising the quality and reliability of the response. 

In a more advanced implementation, the MCP server might expose a dedicated tool such as \texttt{get\_fls\_output()}. This abstraction makes it easier for the LLM to invoke the underlying functionality without constructing the full command manually. However, it still assumes that the LLM understands that \texttt{fls} is the appropriate tool for listing deleted files. A further refinement would be to expose a semantically named tool like \texttt{get\_file\_system\_hierarchy\_using\_fls()}, accompanied by a detailed docstring that provides contextual guidance. This helps the LLM better align the user’s intent with the correct tool. Nevertheless, the model is still responsible for parsing the output and identifying deleted files—typically indicated by an asterisk in the \texttt{fls} output---introducing another potential point of failure.

To further reduce ambiguity and guide the LLM more effectively, the MCP server can expose a high-level tool such as \texttt{list\_deleted\_files\_using\_fls()}. In this case, the LLM’s role is limited to identifying and invoking the appropriate tool based on the user’s query. 

\paragraph{Correlation.}
In addition to enabling access to individual artifact sources or forensic tools, MCP can support the correlation of data across multiple sources. Naturally, one approach is to implement a single MCP server that exposes high-level functions, such as \texttt{list\_deleted\_files()}, which internally aggregate and correlate results from multiple tools or data sources. In such cases, the logic for correlation is hardcoded within the MCP server, offering the LLM a unified and context-aware result. However, this approach limits the LLM’s flexibility, as it has no influence over how data is combined or interpreted.

An alternative strategy leverages the LLM’s ability to interact with multiple independent MCP servers. In this approach, the LLM orchestrates artifact analysis by invoking a combination of tools and querying diverse data sources---such as using TSK, a file carver like Scalpel, and inspecting application-level artifacts, including entries in the trash or recycle bin. This gives the LLM greater inferential freedom during the analysis and correlation but also introduces risks: decisions may lack consistency, and correlations may be incomplete or incorrect.

Ultimately, the effectiveness of MCP in artifact analysis depends on how responsibilities are distributed between the LLM and the server. Properly designed MCP interfaces can significantly reduce hallucinations and errors arising from limited knowledge or incorrect assumptions. However, this comes at the cost of increased implementation complexity and a greater need for forensic domain expertise. As MCP becomes integrated into forensic workflows, practitioners must carefully consider the abstraction level each server provides and strike an appropriate balance between automation, transparency, and control.
\subsection{Comprehensible Reporting}
As highlighted by Michelet and Breitinger~\cite{michelet2024chatgpt}, LLMs show promise in supporting forensic report writing. 
Their analysis of typical forensic reports revealed a common structure based on four main data sources: tool reports generated by forensic software, the prosecutor’s mandate outlining the case and investigative goals, lab logs documenting the examiner’s procedures and observations, and the examiner’s own knowledge drawn from both the current and past cases.

In their approach, LLMs were prompted manually with structured inputs corresponding to these sections. 
While effective, this workflow is limited by its reliance on manual data collation and prompting. 
We propose that integrating LLMs with MCP servers may significantly improve the efficiency, consistency, and traceability of this process.

With MCP integration, the LLM no longer depends on manually aggregated inputs. 
Instead, it can autonomously retrieve and interpret relevant information from specialized MCP servers---each responsible for providing one type of contextual data (e.g., tool outputs, mandate content, lab notes, or prior case knowledge). 
A shared knowledge base guides the LLM in structuring the report according to formal forensic standards and include information, that lets an LLM ask for examiner knowledge or additional data or information when required.

Given a high-level request---such as generating a report for a particular case---the LLM can use MCP interfaces to identify and retrieve query-relevant information, organize it based on a predefined format template, and tailor the content to match the required detail level. 
This enables more dynamic and adaptable reporting workflows, suitable for different forensic use cases.

Furthermore, the MCP servers can log the exact data they provide to the LLM, helping to comprehend the report generated by the LLM. 
Notably, this shows us what kind of data could have helped the LLM to write the report but does not mean that all the retrieved data is included in the report.

%
\subsection{Agentic MCP Server for Live Analysis and Response} \label{sec:agentic-mcp-for-ir}
Endpoint agents are commonly installed on client systems to monitor activity and respond to threats. MCP provides a flexible way to simplify interactions with such agents. For example, an MCP server can be deployed directly as an endpoint agent on a client, exposing core monitoring capabilities to an LLM, such as listing running processes, active network sockets, or logged-in users. This allows the LLM to query live system data that may be critical during incident response scenarios.

Beyond data collection, the MCP server can also expose response actions to the LLM, including terminating user sessions, closing network connections, or killing specific processes. This enables the LLM to make context-aware decisions based on up-to-date system state---for instance, detecting an unauthorized SSH session and terminating it in real time.

The degree of autonomy granted to the LLM can vary significantly. In a controlled setup, the LLM is restricted to a set of predefined, well-documented tools provided by the MCP server. Alternatively, a less restrictive configuration might expose a full root shell, allowing the LLM to execute arbitrary commands on the system. While more powerful, such an approach increases the risk of unintended consequences due to malformed or incorrect instructions. Alternatively, MCP can act as a translation layer between the LLM and existing endpoint agent frameworks. Instead of replacing these established agents, the MCP server can wrap their APIs or command-line interfaces and expose selected functionalities as structured tools.

%
\subsection{Automated Synthetic Data Generation}
A persistent challenge in digital forensics is the automatic generation of test data to train investigators and validate forensic tools. 
As Voigt et al.~\cite{voigtReimagenGeneratingCoherent2024} highlight, LLMs hold significant potential for creating forensic data with minimal instructor effort by leveraging their creative capacity to simulate diverse personas, backgrounds, and actions.

With MCP, servers can interact with tools like ForTrace++~\cite{wolfHypervisorbasedDataSynthesis2024a}, enabling LLMs to generate scenario descriptions in natural language. 
For example, a prompt like ``Create a ForTrace++ scenario where Firefox navigates to example.com'' can be interpreted by the LLM, which---using an MCP server’s knowledge of the scenario format---automatically produces a valid scenario file.

MCP also supports more advanced simulations through dynamic GUI automation. 
When the server provides capabilities such as capturing screenshots, identifying GUI elements, and simulating keyboard and mouse input, the LLM can iteratively observe and interact with the user interface. 
It may request a screenshot, analyze it via computer vision, determine the next action based on the visible UI, and instruct the server to perform mouse or keyboard inputs. 
This allows the LLM to handle unpredictable interface behavior, such as unexpected pop-ups or modal dialogs, with appropriate responses.

Even more complex scenarios can be envisioned where multiple LLMs, each connected to its own MCP server, interact with one another. 
For example, two LLMs running Android emulators could exchange messages, respond to each other’s actions, and collaboratively simulate realistic multi-device, multi-user forensic scenarios.

%

%
\subsection{Adversary Simulation}
MCP can also be employed in proactive security scenarios, such as simulating attacker behavior in controlled environments. By equipping an LLM with tools commonly used throughout the attack lifecycle---such as network scanners, enumeration scripts, credential dumpers, privilege escalation exploits, and lateral movement utilities---the LLM can act as an autonomous or semi-autonomous adversary within a sandboxed lab environment. Exposed to these tools via MCP, and guided by a clear objective (e.g., exfiltrate a file or gain persistence), the LLM can query system state, invoke actions, and adapt its strategy in real time. This offers a novel and dynamic approach to red teaming, penetration testing, and adversary simulation---where the LLM behaves less like a scripted agent and more like a flexible, goal-driven attacker. Recent work has highlighted the need for such intelligent simulation platforms to assess vulnerabilities against AI-powered threats and support defenders in maintaining control over increasingly complex attack surfaces~\cite{guembe2022emerging,jaber2022towards}.

Another interesting direction is to deploy the agent from Section~\ref{sec:agentic-mcp-for-ir} as a defensive counterpart within the simulation environment. This creates the possibility of a closed-loop, ``Colosseum-style'' simulation where offensive and defensive systems adapt to each other in real time, offering a unique testbed for studying both attack strategies and automated incident response dynamics.

In particular, such simulations help uncover detection blind spots by introducing dynamic, LLM-generated attack chains that may deviate from established Tactics, Techniques, and Procedures. Unlike traditional testing approaches that rely on known signatures or scripted behavior, LLM-driven adversaries can combine tools and strategies in novel ways, revealing weaknesses that static or rule-based test cases might overlook. This allows defenders to evaluate their detection capabilities against a broader spectrum of potential threats, including those that do not conform to familiar adversary patterns.

Beyond detection, these simulations also enable the development and validation of automated forensic playbooks---predefined workflows for responding to security incidents. Running LLM-driven attacks in a controlled environment allows security teams to test whether these playbooks trigger appropriately, adapt to evolving threats, and capture the necessary evidence at the right time. This helps identify procedural gaps and refine decision logic under conditions that more closely reflect real-world, unpredictable adversary behavior. As a result, MCP-enabled simulations support both automation and analyst training by offering a more authentic environment for testing and improving incident response.


%
\section{Discussion}
As demonstrated in the previous sections, MCP opens up a broad range of opportunities for digital forensics. These range from simplifying common tasks such as artifact analysis, to enabling more sophisticated operations like cross-source correlation and the autonomous generation of test data. However, to fully leverage MCP’s potential within forensic workflows, several technical and conceptual considerations must first be addressed. Following this, we examine additional challenges that MCP introduces to the field of digital forensics

\paragraph{Enhancing Transparency in LLM-Driven Analysis.}
A central benefit of MCP is its ability to reduce the inherent black-box character of LLMs. When integrated with context-aware, application-specific MCP servers, LLMs no longer need to rely solely on internal reasoning to interpret data or make decisions. Instead, they can invoke clearly defined tools whose behavior is preconfigured and documented. This improves transparency and makes the analysis process more understandable and traceable---both for analysts and for future reviews.

To ensure this level of transparency, MCP servers should offer detailed documentation and implement comprehensive logging. Logs should capture each tool invocation, including parameters passed by the LLM, tool output, and timestamps. When paired with LLM logs and structured explanations of the tools used, the auditability of automated forensic workflows increases significantly.

\paragraph{Delegating Microtasks to Custom MCP Servers.}
MCP’s design allows forensic workflows to be broken down into smaller, well-defined components. Even micro-level decisions that would typically be made implicitly by the LLM---such as filtering data or normalizing timestamps---can be delegated to lightweight MCP tools. This modularization increases reproducibility and clarity, as each step in the analysis becomes externally visible and testable.

\paragraph{Standardization of MCP Servers.}
To support the reliable integration of MCP servers into forensic workflows, it is essential to establish standardized definitions for the tools and functions they expose. For example, a function named \texttt{extract\_deleted\_files()} is ambiguous, as it does not specify the method of extraction—such as file carving, recovery from a trash bin, or metadata-based reconstruction. Leveraging existing frameworks like SOLVE-IT~\cite{hargreavesSOLVEITProposedDigital2025} can help define consistent tool semantics and documentation standards, facilitating the collaborative development and reuse of MCP tools among forensic practitioners.

\paragraph{Ensuring Forensic Soundness and Data Integrity.}
For MCP to be usable in evidentiary contexts, its implementation must follow established principles of forensic soundness. This includes ensuring that tools accessed by the LLM are strictly read-only and do not modify the underlying data. The MCP server itself must also operate non-intrusively, meaning it should not alter, overwrite, or delete any evidence---intentionally or unintentionally---unless such actions are explicitly permitted in a controlled and well-documented context. Maintaining this level of control is vital to preserve the integrity of the evidence and to ensure that all results remain verifiable and legally defensible. Existing general-purpose MCP servers for database access often include both read and write functionality and may, therefore, be unsuited for forensic use without modification.

\paragraph{Facilitating Privacy through Pseudonymization.}
Another challenge in digital forensics---especially when relying on hosted LLMs---is the handling of sensitive or personal data. MCP can serve as a useful abstraction layer here as well. Servers can be designed to pseudonymize sensitive identifiers before exposing data to the LLM, thus mitigating privacy concerns during analysis.

By integrating pseudonymization functionality into the MCP server, identifiers and other sensitive values can be consistently replaced before being passed to the LLM. This requires case-specific design to ensure reversibility (for de-pseudonymization) and correctness. Still, it presents a promising path to enabling the use of powerful hosted LLMs without exposing sensitive case material.

\paragraph{Auditing MCP Servers.}
A thorough audit must be performed beforehand to integrate an MCP server into forensic workflows safely and securely. Even if an MCP server fundamentally implements the aforementioned beneficial features, such as logging, it must be ensured that the intended functionality has indeed been correctly implemented. In particular, care must be taken to ensure no unintended or even malicious functionality is present in the MCP server. For example, it would be entirely feasible to develop an MCP server that operates correctly and possesses all the necessary forensic features yet additionally transmits certain sensitive information from case data to the Internet. To counter this issue, auditing and analyzing MCP servers is required. Trust can be further enhanced by maintaining a curated repository that exclusively hosts reviewed and accredited MCP servers.

\paragraph{Forensic Analysis of MCP Servers.}In addition to the analysis conducted before the integration of MCP servers into forensic workflows, the analysis of MCP servers will also play a role in future investigations. Integrating these servers into operating systems and applications will inevitably lead to their presence while examining IT systems. Consequently, there is an equally strong need for capabilities to analyze MCP servers forensically. This includes, for example, determining its functionalities, analyzing function executions and their associated input and output data, and the identification of other potentially present artifacts generated within or by an MCP server.

\paragraph{Action Attribution.}
Beyond the opportunities that MCP offers for digital forensics and incident response, it also adds a new layer to an already complex challenge: the attribution of actions. Determining the origin of an action---whether it was initiated by a user, a piece of malware, or an automated system---has long been a core task in forensic investigations~\cite{bowles_hernandez-castro.trojan-horse-defence.2015}. The integration of MCP introduces a new dimension to this challenge: actions can now be executed by MCP servers, potentially triggered by an autonomous AI agent on the user's computer acting independently or on behalf of users. As MCP becomes increasingly embedded into applications and operating systems, enabling agents to control fundamental system functionality, forensic investigations must now account for AI-driven behavior as a distinct source of activity.

Consider a scenario in which a web browser visits a website and downloads illicit content. Traditionally, a forensic investigation would focus on identifying artifacts that support the hypothesis that the user performed these actions---such as web history entries, download records, or file system traces. These artifacts could be found across browser databases, file metadata, and system logs.

Now, imagine a system where an autonomous AI agent, connected via MCP, has full control over the browser and is capable of initiating visits and downloads independently. In this case, attribution becomes substantially more difficult. The investigator must establish whether the user acted intentionally or whether the AI performed these actions autonomously---potentially opening the door to defenses such as the Trojan Horse Defense, in which a suspect claims the system acted on its own without their knowledge.

To address this, it will become critical to identify forensic artifacts that can reliably distinguish between actions carried out by a human user and those initiated by an AI agent. This may involve analyzing MCP server logs, agent execution traces, or internal mechanisms of the controlled applications. Moreover, the notion of \textit{intent} and \textit{human involvement} introduces an additional layer of complexity. If an AI agent executed a task directly but was explicitly instructed to do so by the user, the responsibility differs significantly from a case where the agent acted autonomously or in error. Future investigations will not only need to determine what happened, but also who caused it to happen, and in what context---human, AI, or a combination of both.

%
\section{Conclusion}
MCP demonstrates significant potential to enhance the forensic applicability of LLMs by addressing core challenges such as transparency, explainability, and reproducibility. 
In this paper, we propose a conceptual framework that demonstrates how MCP servers can constrain the interpretive scope of LLMs in a structured and auditable manner, thereby helping to align LLM-supported analysis with forensic expectations.
Using a theoretical exploration of relevant use cases, we discussed how MCP can serve as a basis for more transparent and modular workflows. 
Our discussion also identified critical considerations, such as forensic soundness and privacy protection, that must guide the design and evaluation of MCP-based systems to ensure their suitability in investigative settings.

Despite its potential, applying MCP in forensic contexts presents some limitations. 
While MCP servers can provide contextual data to the LLM, it remains uncertain whether the model reliably incorporates this information into its reasoning or output. Combined with the ongoing risk of hallucinations, this limitation should be examined more closely in future research. 
Furthermore, MCP servers may themselves rely on complex internal logic or even external model calls, which---if not properly documented---can reintroduce black-box characteristics that undermine transparency.
Additionally, insufficient testing or lack of standardization across MCP server implementations may lead to inconsistencies that affect reproducibility and admissibility in court.

Building on our theoretical foundation, future research should explore practical strategies for integrating MCP servers into existing forensic tools and workflows. A key focus for this will be determining how responsibilities should be divided between the MCP server and the underlying LLM to ensure transparency, auditability, and reproducibility.

Additionally, as LLMs become increasingly embedded in everyday applications, new forensic challenges will emerge. MCP logs may aid in reconstructing user behavior, however, the semi-autonomous nature of LLMs complicates efforts to attribute specific actions to either the user or the LLM. Resolving this ambiguity will be critical to maintaining forensic interpretations that are grounded, reliable, and consistent with legal standards.

\bibliographystyle{splncs04}
\bibliography{main.bbl}
%




\end{document}